\documentclass[
    ,final            
  ]{aipproc}

\layoutstyle{6x9}
\usepackage{amsmath}
\newcommand\beq{\begin{equation}}
\newcommand\eeq{\end{equation}}
\newcommand\beqa{\begin{eqnarray}}
\newcommand\eeqa{\end{eqnarray}}

\newcommand\bzeta{\mbox{\boldmath$\zeta$}}

\newcommand\bsigma{\mbox{\boldmath$\sigma$}}

\newcommand\bk{{\bf k}}
\newcommand\bq{{\bf q}}

\newcommand\br{{\bf r}}
\usepackage{graphicx}
\begin{document}

\title{Microscopic origin of the magnetic field in compact stars}


\author{Toshitaka Tatsumi}{
  address={Department of Physics, Kyoto University, Kyoto 606-8502, Japan}}



\begin{abstract}
A magnetic aspect of quark matter is studied by the Fermi liquid
 theory. 
The magnetic susceptibility is
 derived  with the one-gluon-exchange interaction, and the critical Fermi
 momentum for spontaneous spin polarization is found to be
 $1.4$fm$^{-1}$.
A scenario about the origin of magnetic field in compact stars is
 presented by using this result. 
\end{abstract}

\maketitle


\section{Introduction}

Nowadays the phase diagram of QCD in the density-temperature plane has
been explored by many people: the high-temperature region is relevant
for relativistic heavy-ion collisions or cosmological phase transition in
early universe, while the high-density region for compact stars. In the
high-density and low-temperature region many exciting phenomena have
been expected due to the large and sharp Fermi surface, such as color
superconductivity, chiral density waves or ferromagnetism. Here we are
concentrated in the magnetic aspect of quark matter: we are interested
in the spin degree of freedom \cite{rev05}. If magnetism is realized in quark matter,
it should be interesting theoretically, but have important implications
for the magnetic properties of compact stars.

Recent discoveries of magnetars, compact stars with huge magnetic field
of $O(10^{15}{\rm G})$, seem to enforce us to reconsider the origin of
the magnetic field in compact stars. They have firstly observed by the
$P-{\dot P}$ curve, and some cyclotron absorption lines have been
recently 
observed \cite{woo}. A naive working hypothesis of
conservation of magnetic flux during their evolution from the
main-sequence progenitors cannot be applied to magnetars, because the
resultant radius is too small for $O(10^{15}{\rm G})$. The dynamo
mechanism may work in compact stars, but it might look to be unnatural to
produce such a huge magnetic field. The typical energy scale is given by
the interaction energy with this magnetic field, which amounts to O(MeV)
for electrons, and O(keV) to O(MeV) for nucleons or quarks. On the other
hand, considering
the atomic energy scale or the strong-interaction energy scale, 
we can say that $O(10^{15}{\rm
G})$ is very large for electrons, but not so large for nucleons or
quarks.
Hence it would be interesting to consider a microscopic origin of
magnetic field as an alternative: if ferromagnetism or spontaneous spin
polarization occurs inside compact stars, it may be a possible
candidate
\footnote{In a recent paper Makishima also suggested a hadronic origin
from the observation of X-ray binaries \cite{mak}.}
. For nuclear matter, there
have been done many calculations with different nuclear
forces and different methods since the first
discoveries of pulsars in early seventies, but all of them have given 
negative results so far \cite{fan01}. In the following we discuss the
magnetic aspect of quark matter and consider a possibility of
spontaneous spin polarization.

\section{Relativistic ferromagnetism}

We have considered the possibility of
ferromagnetism in quark matter interacting with the one-gluon-exchange
(OGE) 
interaction \cite{tat00} or with an effective interaction \cite{nak03}, and
suggested that quark matter has a potentiality to be spontaneously polarized.
To understand the magnetic properties of quark matter 
more realistically , especially near the critical point, 
some non-perturbative consideration about the instability of the
Fermi surface is indispensable.  Recently there are some studies about
the effective interaction near the Fermi surface \cite{hon,han}, using the idea
of the renormalization group \cite{pol}.
Here we apply the Fermi liquid theory
to derive the magnetic susceptibility and discuss the spontaneous spin
polarization, considering quarks as quasiparticles \cite{bay,her,neg}.

\subsection{Fermi liquid theory}

In the Fermi liquid theory the total energy is given as a functional of
the distribution function. 
As is already shown in ref.~\cite{mar01}, the spin degree of freedom is
specified by the three vector $\bzeta$ in the rest frame.
Then the quasi-particle energy and the effective interaction near the Fermi
surface can be written as,
\beq
\epsilon(\bk\zeta ci)=\frac{\delta E}{\delta n(\bk\zeta ci)},~~~
f_{\bk\zeta c i,\bq\zeta' d j}=\frac{\delta\epsilon(\bk\zeta c i)}{\delta
n(\bq\zeta' d j)},
\eeq
where the subscripts $c$($d$) and $i$($j$) denotes the color and flavor
degrees of freedom. The Landau Fermi liquid interaction $f_{\bk\zeta c
i,\bq\zeta' d j}$ is related to the forward scattering amplitude for two
quarks on the Fermi surface.
\footnote{Note that it is also only the non-relevant interaction at the
Fermi surface in the context of the renormalization group approach \cite{pol}.}
In QCD the interaction is flavor independent, 
$f_{\bk\zeta c i,\bq\zeta' d j}=\delta_{ij}f_{\bk\zeta c,\bq\zeta' d}$.
Since there is no direct
interaction due to color neutrality, the Fock exchange
interaction gives the leading contribution in the weak coupling limit, i.e. the color symmetric OGE
interaction can be written as
\beq
f^S_{\bk\zeta,\bq\zeta'}\equiv\frac{1}{N_c^2}\sum_{c,d}f_{\bk\zeta c,\bq\zeta'
d}=\frac{m}{E_k}\frac{m}{E_q}M_{\bk\zeta,\bq\zeta'},
\label{fint}
\eeq 
with the Lorentz invariant matrix element,
\beq
M_{\bk\zeta,\bq\zeta'}=g^2\frac{N_c^2-1}{4N_c^2m^2}\left[2m^2-k\cdot q-m^2a\cdot b\right]\frac{1}{(k-q)^2},
\label{invm}
\eeq
where we used the Feynman gauge for the gluon propagator. The term
including the inner product $a\cdot b$ represents the spin
dependence. The spin vector $a^\mu$ is explicitly given as a function of $\bzeta$ and
momentum \cite{mar01}. There are many possible forms about $a^\mu$, but we
here use the simplest one,
\beq
a^0=\frac{\bk\cdot\bzeta}{m}, {\bf a}=\bzeta+\frac{\bk(\bzeta\cdot\bk)}{m(E_k+m)}.
\label{sta}
\eeq 

From the invariance of the properties of quark matter under the Lorentz
transformation, the Fermi velocity can be written \cite{bay} as
\beq
v_F^{-1}=\left(\frac{\partial k}{\partial\epsilon(\bk\zeta)}\right)_{k_F}=\frac{\mu}{k_F}\left(1+\frac{1}{3}F_1^S\right)
\label{fvel}
\eeq
with the spin-symmetric Landau parameter $F_1^S$ defined by
\beq
F_1^S=N(0)f_1^S,
f_1^S=-\frac{3}{4}\frac{g^2(N_c^2-1)}{4N_c^2\mu^2}\frac{m^2}{k_F^2}\int_{-1}^{1}duu\frac{1}{1-u}, 
\eeq
for OGE.
Here $N(0)$ is the density of states at the Fermi surface,
$N(0)=2N_ck_F^2/2\pi^2(\partial k/\partial\epsilon(\bk\zeta)_{k_F})$. 
Note that $f_1^S$ clearly shows log divergence reflecting the gauge interaction. When we take into account the
higher-order corrections for the gluon propagator, the electric
propagator is screened by the Debye mass, while the magnetic one
receives only the Landau damping. This fact exhibits the non Fermi
liquid nature of quark matter. However, we shall see that the magnetic
susceptibility becomes finite even in this case.

Applying the weak magnetic field to quark matter, we consider the energy
change (see Fig.~1).  
\begin{figure}[h]
  \includegraphics[height=.2\textheight]{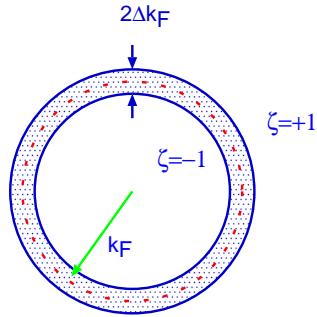}
  \caption{Modification of the Fermi surface in the presence of the weak
 magnetic field.}
\end{figure}
Using the Gordon identity, the QED interaction Lagrangian can be recast
as
\beqa
\int d^4x{\cal L}_{\rm ext}^{\rm QED}
=\sum_f\mu_q^f\int d^4x{\bar\psi}_f\left[-i\br\times\nabla+\bsigma\right]\times{\bf B}\psi_f
\eeqa 
with the magnetic moment, $\mu_q^f=e_q^f/2m$.
Since the orbital angular momentum gives null contribution on average,
we hereafter only consider the spin contribution. In the following we
consider only one flavor without loss of generality. For the energy to be
minimum (chemical equilibrium),
\beq
\epsilon(k_F+\Delta k_F, \zeta=+1)=\epsilon(k_F-\Delta k_F, \zeta=-1),
\eeq
i.e.,
\beqa
&-&\frac{g_D\mu_q B}{2}+\left(\frac{d\epsilon}{dk}\right)_{k_F}\Delta k_F+({\bar
f}_{++}-{\bar f}_{+-})\Delta N\nonumber\\
&=&\frac{g_D\mu_q B}{2}-\left(\frac{d\epsilon}{dk}\right)_{k_F}\Delta k_F+({\bar
f}_{-+}-{\bar f}_{--})\Delta N,
\label{equil}
\eeqa
where $\Delta N=N_ck_F^2\Delta k_F/2\pi^2$, and $g_D$ is the
gyromagnetic ratio \cite{mar01},
\beq
g_D=2\int\frac{d\Omega_k}{4\pi}\left(a_z-\frac{k_F}{\mu}\cos\theta a_0\right),
\eeq
which is reduced to be 2 in the non-relativistic limit, $m\gg k_F$. The
angle-averaged Fermi liquid interactions ${\bar f}_{\zeta\zeta'}$ are
given as
\beqa
{\bar f}_{++}={\bar
f}_{--}&=&\frac{(N_c^2-1)g^2}{4N_c^2\mu^2}\left[\frac{1}{2}-\frac{m^2}{k_F^2}
-\frac{1}{3}\frac{m(\mu-m)}{k_F^2}\right]+\frac{f_1^S}{3}\nonumber\\
{\bar f}_{+-}={\bar f}_{-+}&=&\frac{(N_c^2-1)g^2}{4N_c^2\mu^2}\left[\frac{1}{2}+\frac{1}{3}\frac{m(\mu-m)}{k_F^2}\right],
\label{fintav}
\eeqa
by the use of the standard spin configuration (\ref{sta}).
The latter is reduced to null in the non-relativistic limit, which
implies there is no interaction between quarks with different spins
\cite{her}. From Eqs.~(\ref{equil}) and (\ref{fintav}) we find the spin
susceptibility,
\beqa
\chi_{\rm spin}&\equiv& g_D\mu_q\Delta N/VB\nonumber\\
&=&\chi_{\rm free}\left[1-\frac{4\alpha_c}{3\mu\pi}\frac{m(2\mu+m)}{3k_F}\right]^{-1}
\eeqa
for $N_c=3$, with the corresponding one without any interaction, $\chi_{\rm
free}=g^2_D\mu_q^2\mu k_F/4\pi^2$. Note that log divergence is included
in the Fermi liquid interaction (\ref{fintav}), but it is canceled by the
one coming from the Fermi velocity (\ref{fvel}).
Fig.~\ref{sus} shows the ratio of spin
susceptibility as a function of the Fermi momentum. It diverges around
$k_F=1.4$fm$^{-1}$, which is a signal of the spontaneous magnetization.

\begin{figure}[h]
  \includegraphics[height=.2\textheight]{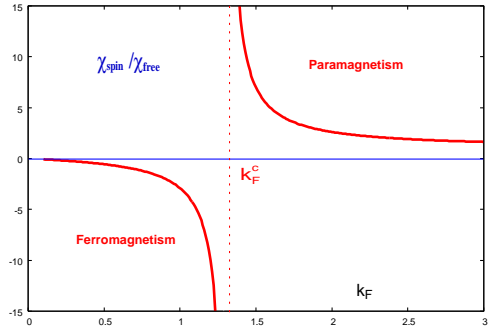}
  \caption{Spin susceptibility as a function of the Fermi momentum.}
\label{sus}
\end{figure}

It would be interesting to compare the above result with the previous
one given by the perturbative calculation with OGE \cite{tat00}, where
we can see the {\it weakly first-order} phase transition at a certain low
density, and that 
the critical density is similar to that given by the
Fermi liquid theory. Thus two calculations are consistent with each
other in the weak coupling limit.

Here we have discussed only the lowest order contribution, but a recent
paper has also suggested the phase transition at low densities by
including the higher-order effects \cite{nie}.

\section{Astrophysical implications}

There are two possibilities about the existence of quark matter in
compact stars: one is in the quark stars, which are composed of 
low density strange matter, and the other is in the core region of
neutron stars, which are called hybrid stars. Consider magnetars as
quark stars or hybrid stars. Then we can easily estimate their magnetic
field near the surface, if ferromagnetism is realized in quark matter.
The maximum strength of the magnetic field on the surface $r=R$ is
estimated by 
\beq
B_{\rm max}=\frac{8\pi}{3}\left(\frac{r_Q}{R}\right)^3\mu_Qn_Q,
\eeq
where $r_Q$ is the radius of quark lump with density $n_Q$ and $\mu_Q$
the single quark magnetic
moment; 
it amounts to 
$O(10^{15-17})$G for $n_Q=0.1$fm$^{-3}$, which might be enough for
magnetars.
It should be interesting to observe the braking indexes about 
magnetars; if their values are near three, the dipole
radiation is a good picture and we may say that the above scenario looks more realistic.

There may be left another interesting problem about hierarchy of magnetic
field in compact stars (Table 1). Unfortunately, the idea of ferromagnetism may not be 
sufficient for explaining it, and we need to consider the global magnetic structure
and some dynamical mechanisms,
e.g. formation of magnetic domain or existence of metamagnetism, besides it.

\begin{table}[h]
\begin{tabular}{|c||c|c|c|}\hline
&millisecond pulsars & usual radio pulsars & magnetars \\\hline\hline
Magnetic field [G] & $10^9$ &$10^{12}$ &$10^{15}$\\
Period [sec] & $10^{-2}$ & $10^0$ & $10^1$ \\
Age [year] & $10^9$ & $10^6$ & $10^3$ \\\hline
\end{tabular}
\caption{Hierarchy of magnetic field in compact stars.}
\end{table}

We might also consider a scenario about the cosmological magnetic field in
the galaxies and extra galaxies. It is well known that magnetic fields
are present in all galaxies and galaxy clusters, which are characterized
by the strength, $10^{-7}-10^{-5}$G, with the spatial scale, $\leq 1$Mpc
\cite{wid}. The origin of such magnetic fields is still unknown, but the
first magnetic fields may have been created in the early universe. If
magnetized quark lumps are generated during the QCD phase transition,
they can give the seed fields.

\section{Concluding remarks}

Magnetic properties of quark matter have been discussed by applying the
Fermi liquid theory, which gives one of the non-perturbative tool to
analyze them. Spin couples with motion in the relativistic theories, and
we must extend the Fermi liquid interaction accordingly. 
We have seen that the magnetic susceptibility can be given
by applying a weak magnetic field and considering the energy change in
the tiny region near the Fermi surface. The Landau parameters may be log
divergent for the gauge interaction, but the magnetic susceptibility
is given to be finite by the cancellation. The critical point is the
found to be the same order with the one given by the perturbative
evaluation, which may support the relevance of the Fermi liquid theory
in this problem. We may extend the present analysis by including higher
order diagrams within the Fermi liquid theory or using the
renormalization group.

If ferromagnetism is realized in quark matter, we may consider various
scenarios: it might give a microscopic origin of the magnetic field
in compact stars, and a seed of the primordial magnetic field during the
cosmological QCD phase transition. It
may also predict the production of small magnets composed of strange
matter during the relativistic heavy-ion collisions. 

  This work is supported by the Japanese Grant-in-Aid for Scientific
Research Fund of the Ministry of Education, Culture, Sports, Science and
Technology (13640282,16540246).




\end{document}